\documentclass[onecolumn,showpacs,11pt]{revtex4}
\usepackage{graphicx}
\usepackage{dcolumn}
\usepackage{bm}
\usepackage{color}
\begin{document}
\newcommand{\hs}{\hspace*{0.5cm}}
\newcommand{\vs}{\vspace*{0.5cm}}
\newcommand{\be}{\begin{equation}}
\newcommand{\ee}{\end{equation}}
\newcommand{\bea}{\begin{eqnarray}}
\newcommand{\eea}{\end{eqnarray}}
\newcommand{\ben}{\begin{enumerate}}
\newcommand{\een}{\end{enumerate}}
\newcommand{\bde}{\begin{widetext}}
\newcommand{\ede}{\end{widetext}}
\newcommand{\nn}{\nonumber}
\newcommand{\crn}{\nonumber \\}
\newcommand{\Tr}{\mathrm{Tr}}
\newcommand{\non}{\nonumber}
\newcommand{\noi}{\noindent}
\newcommand{\al}{\alpha}
\newcommand{\la}{\lambda}
\newcommand{\bet}{\beta}
\newcommand{\ga}{\gamma}
\newcommand{\va}{\varphi}
\newcommand{\om}{\omega}
\newcommand{\pa}{\partial}
\newcommand{\+}{\dagger}
\newcommand{\fr}{\frac}
\newcommand{\bc}{\begin{center}}
\newcommand{\ec}{\end{center}}
\newcommand{\Ga}{\Gamma}
\newcommand{\de}{\delta}
\newcommand{\De}{\Delta}
\newcommand{\ep}{\epsilon}
\newcommand{\varep}{\varepsilon}
\newcommand{\ka}{\kappa}
\newcommand{\La}{\Lambda}
\newcommand{\si}{\sigma}
\newcommand{\Si}{\Sigma}
\newcommand{\ta}{\tau}
\newcommand{\up}{\upsilon}
\newcommand{\Up}{\Upsilon}
\newcommand{\ze}{\zeta}
\newcommand{\ps}{\psi}
\newcommand{\Ps}{\Psi}
\newcommand{\ph}{\phi}
\newcommand{\vph}{\varphi}
\newcommand{\Ph}{\Phi}
\newcommand{\Om}{\Omega}
\newcommand{\Vien}[1]{{\color{red}#1}}
\newcommand{\Long}[1]{{\color{green}#1}}
\newcommand{\Antonio}[1]{{\color{blue}#1}}

\title{Comment on the article by D. Borah and B. Karmakar "Linear seesaw for Dirac neutrinos with $A_4$ flavour symmetry", Phys. Lett. B\textbf{789} (2019) 59–70, arXiv: 1806.10685}
\author{V. V. Vien$^{a,b}$}
\email{vovanvien@tdtu.edu.vn}
\author{H. N. Long$^{c}$}
\email{hnlong@iop.vast.ac.vn}
\author{A. E. C\'arcamo Hern\'andez${}^{d}$}
\email{antonio.carcamo@usm.cl}
\affiliation{$^{a}$ Theoretical Particle Physics and Cosmology Research Group, Advanced Institute of Materials Science,
Ton Duc Thang University, Ho Chi Minh City, Vietnam\\
$^{b}$ Faculty of Applied Sciences, Ton Duc Thang University, Ho Chi Minh City, Vietnam.\\
$^{c}$Institute of Physics, Vietnam Academy of Science and Technology, 10
Dao Tan, Ba Dinh, Hanoi Vietnam\\
$^{d}$Universidad T\'{e}cnica Federico Santa Mar\'{\i}a and Centro Cient%
\'{\i}fico-Tecnol\'{o}gico de Valpara\'{\i}so, \\
Casilla 110-V, Valpara\'{\i}so, Chile.}

\begin{abstract}
D. Borah and B. Karmakar \cite{A4linear} have proposed an $A_4$ flavoured linear seesaw model to realise light Dirac neutrinos.
In this comment article, we show that some neutrino Yukawa interactions were missed in the model Ref. \cite{A4linear}, thus implying that a different formula would be needed to determine the effective neutrino mass matrix, with significantly different results. Our result shows that, unlike stated in Ref. \cite{A4linear}, that the inverted neutrino mass spectrum is not ruled out.
\end{abstract}
\date{\today}
\maketitle
A recent study made by D. Borah and B. Karmakar \cite{A4linear} has proposed an $A_4$ flavoured linear seesaw model to realise light Dirac neutrinos. That model predicts in terms of neutrino mass hierarchy, the leptonic Dirac CP phase, octant of atmospheric mixing
angle as well as absolute neutrino masses.

In the $A_4$ flavoured linear seesaw model of Ref. \cite{A4linear}, the Yukawa terms $\bar{N_L} N_R (y_{\xi_3}\xi^+ + y_{s_3}\phi^+_S+y_{a_3} \phi^+_S)$ and $\bar{S_L} S_R
(y_{\xi_4}\xi^{\dagger}+ y_{s_4}\phi^{\dagger}_S+y_{a_4}\phi^{\dagger}_S)$ must exist in the neutrino sector since they are invariant under all symmetries of the model.
Therefore, the two corresponding contributions have to be added in Eq. (22) of Ref. \cite{A4linear} and  the mixing between
the heavy neutrinos $N_LN_R$ and $S_LS_R$ are generated at renormalizable level, with the corresponding mass matrices $M_{NN}, M_{SS}$ given by:
\begin{eqnarray}\label{mat:heavy}
 M_{ N N} =\left(
\begin{array}{ccc}
 x_3 & 0   & s_3+a_3 \\
 0  & x_3 & 0\\
 s_3 - a_3 & 0 & x_3
\end{array}
\right), ~~ M_{ S S} =\left(
\begin{array}{ccc}
 x_4 & 0   & s_4+a_4 \\
 0  & x_4 & 0\\
 s_4 - a_4 & 0 & x_4
\end{array}
\right), \Vien{\label{v2}}
\end{eqnarray}
where \bea
x_3&=&y_{\xi_3} v_{\xi}, \, s_3=y_{s_3} v_{s}, \, a_3=y_{a_3}  v_{s}, \,
x_4 = y_{\xi_4} v_{\xi}, \, 
s_4=y_{s_4} v_{s}, \, a_4=y_{a_4}  v_{s}. \label{xsa} \eea
 In this case, the full neutrino mass matrix has the following form
  \bea
 M_{\mathrm{eff}}&=&  \left(%
 \begin{array}{ccc}
  0&\,\,\, m_{\nu N} &\,\,\, M_{\nu S} \\
 m'_{\nu N} & M_{NN} &\,\,\, M_{NS} \\
 M'_{\nu S} & \,\,\, M'_{NS}  &M_{SS} \\
\end{array}%
\right)\equiv  \left(%
 \begin{array}{ccc}
  0&\,\,\, M_D \\
 M^T_D & M_R \\
\end{array}%
\right), \label{V01} \\
M_D&=&(m_{\nu N} \,\,\, M_{\nu S}),\,\, M^T_D= \left(%
 \begin{array}{c}
  m'_{\nu N}\\
 M'_{\nu S} \\
\end{array}%
\right), \,\, M_R= \left(%
 \begin{array}{cc}
M_{NN} & M_{NS} \\
M'_{NS}  & M_{SS} \\
\end{array}%
\right), \label{V02}
   \eea
which implies that the light Dirac neutrinos mass matrix can be written as \cite{Meff}:
   \bea
   m_{\nu}&=& - M_D M^{-1}_R M^T_D=\left(%
\begin{array}{ccc}
 A_1+A_2 & 0 & B_1+B_2 \\
 0              & A &      0  \\
 B_3+B_4 & 0 &A_3+A_4\\
\end{array}%
\right), \label{mnu}
\eea
where the different entries are given by:
\bea
&& A=(a' b x_1 + a b' x_2 - b b' x_3 - a a' x_4)/(x_1 x_2-x_3 x_4),\crn
&&A_1=(a' b x_1 + a b' x_2 - b b' x_3 -
   a a' x_4) \left[(a_1 - s_1) (a_2 + s_2) - (a_3 - s_3) (a_4 + s_4) \right.\crn
&&\left.  \hs - x_1 x_2 + x_3 x_4\right]/\al,\crn
&&A_2 =\left[a a' (a_4 + s_4)-a' b (a_1 + s_1) + a b' (a_2 + s_2) - b b' (a_3 + s_3)\right]\times \left[(a_2 - s_2) x_1 \right.\crn
&&\left.  \hs +  (a_1- s_1)x_2-( a_4- s_4) x_3 -  ( a_3- s_3) x_4\right]/\al,\crn
&& A_3=(a' b x_1 + a b' x_2 - b b' x_3 - a a' x_4) \left[(a_1 + s_1) (a_2 - s_2) - (a_3 + s_3) (a_4 - s_4) \right.\crn
&&\left.  \hs - x_1 x_2 +x_3 x_4\right]/\al,\crn
&&A_4=\left[b (-a_1 a' + a_3 b' + a' s_1 - b' s_3) + a (a_4 a' - a_2 b' + b' s_2 - a' s_4)\right]\crn
&&\hs \times  \left[(a_2 + s_2) x_1 + (a_1 +
      s_1) x_2 - (a_4 + s_4) x_3 - (a_3 + s_3) x_4\right]/\al ,\crn
&& B_1=\left[a' b (a_1 + s_1) + a b' (a_2 + s_2) - b b' (a_3 + s_3) -
   a a' (a_4 + s_4)\right]\times \left[(a_1 + s_1) (a_2 - s_2)  \right.\crn
&&\left.  \hs + (a_3 + s_3) (-a_4 + s_4) -
   x_1 x_2 + x_3 x_4\right]/\al,\crn
   && B_2=(a' b x_1 + a b' x_2 - b b' x_3 - a a' x_4) \left[(a_2 + s_2) x_1 + (a_1 + s_1) x_2 - (a_4 + s_4) x_3 \right.\crn
&&\left.  \hs- (a_3 + s_3) x_4\right]/\al, \label{A1234B12} \\
&& \al=(a_2 x_1 - s_2 x_1 + a_1 x_2 - s_1 x_2 - a_4 x_3 + s_4 x_3 - a_3 x_4 +
     s_3 x_4)\crn
     &&\hs \times \left[(a_2 + s_2) x_1 + (a_1 + s_1) x_2 - (a_4 + s_4) x_3 - (a_3 + s_3) x_4\right] \crn
     &&\hs +\left[(a_1 + s_1) (a_2 - s_2) + (a_3 + s_3) (-a_4 + s_4) - x_1 x_2 +
     x_3 x_4\right] \crn
     && \hs \times \left[(a_1 - s_1) (a_2 + s_2) + (-a_3 + s_3) (a_4 + s_4) - x_1 x_2 +
    x_3 x_4\right]. \label{alpha}
\eea
Here $m_\nu =a \textbf{I}$, $m'_\nu =a' \textbf{I}$, $M_{\nu S} =b \textbf{I}$, $M'_{\nu S} =b' \textbf{I}$ and $x_{3,4}, a_{3,4}, s_{3,4}$ are given by Eq. (\ref{xsa}) and all the other parameters (e.g, $x_{1,2}, a_{1,2}, s_{1,2}$,...) are the same as in Ref. \cite{A4linear}.

To diagonalise the light active neutrino mass matrix $m_{\nu}$ in Eq. (\ref{mnu}), we define a Hermitian matrix $\mathcal{M}$, given by
\bea
\mathcal{M}&=& m^+_{\nu}m_{\nu}
=  \left(%
\begin{array}{ccc}
 m_{11}&0 &m_{13}  \\
 0          &m_{22} & 0  \\
m^*_{13}& 0 &m_{33}\\
\end{array}%
\right),\label{Unu}
\eea
where
\bea
m_{11}&=&(A_1 + A_2) (A^*_1 + A^*_2) + (B_3 + B_4) (B^*_3 + B^*_4),\crn
m_{22}&=& |A|^2, \crn
m_{33}&=&(A_3 + A_4) (A^*_3 + A^*_4) + (B_1 + B_2) (B^*_1 + B^*_2),\crn
m_{13}&=&(A^*_1 + A^*_2) (B_1 + B_2) + (A_3 + A_4) (B^*_3 + B^*_4), \label{mij}
\eea
with $A, A_{1,2,3,4}$ and $B_{12}$ are defined in Eqs. (\ref{A1234B12}) and (\ref{alpha}).

The mass matrix $\mathcal{M}$ is diagonalized by the rotation matrix $U_{13}$ given by Eq. (21) of Ref. \cite{A4linear}
and the light active neutrino masses $m^2_{1,2,3}$ are given by
\bea
 m^2_{1,3}& =&\frac{1}{2}\left(m_{11}+m_{33}\pm \sqrt{(m_{11}-m_{33})^2+4 |m_{13}|^2}\right)\equiv A+B, \crn
 m^2_2&=&m_{22}.  \label{m123}
 \eea
 We see that the leptonic mixing matrix in this case is the same as in Ref.\cite{A4linear}, however, the expressions of neutrino masses are different from each other. For instance, in the case of normal neutrino mass hierarchy, by taking the best fit values on neutrino mass squared differences given in Ref.\cite{PDG2018}, $\Delta m_{21}^{2}=m^2_2-m^2_1=7.53\times 10^{-5} \mathrm{eV}^2$ and $\Delta m_{32}^{2}=m^2_3-m^2_2=2.51\times 10^{-3}\mathrm{eV}^2$, we find the solution: $ A=0.00121735\, \mathrm{eV}+m_{22}, \,\,
B=- 0.00129265\,\mathrm{eV}.$

 The absolute values of neutrino masses as well as the neutrino mass ordering is still unknown, however, we can use the neutrino oscillation experimental data for normal hierarchy given in Ref. \cite{PDG2018} to find: $m_2\sim  0.0087\, \mathrm{eV}$. In this case, the parameters $A, B$ and the other neutrino masses are explicitly
 given as $A=1.29304\times 10^{-3} \, \mathrm{eV}, B=-1.29265\times 10^{-3} \, \mathrm{eV}$, $m_1=6.245\times 10^{-4}\, \mathrm{eV},\, m_3=5.08497\times 10^{-2}\, \mathrm{eV}$. Then, the resulting sum of neutrino masses takes the form $\sum m_\nu \sim 0.06017 \, \mathrm{eV}$, which is consistent with the current cosmological constraints $m_{\nu}< 0.17 $ eV given in Ref. \cite{constraint17}.

 On the other hand, in the inverted neutrino mass spectrum \cite{PDG2018}, $\Delta m_{21}^{2}=m^2_2-m^2_1=7.53\times 10^{-5} \mathrm{eV}^2$ and $\Delta m_{32}^{2}=m^2_3-m^2_2=-2.56\times 10^{-3}\mathrm{eV}^2$, we obtain: $A=-0.00131765\, \mathrm{eV} + m_{22}, \,\,
B=0.00124235\, \mathrm{eV}$. Then, in the scenario of inverted neutrino mass ordering, we find $m_2=0.051\, \mathrm{eV}$, with the parameters $A, B$ and the other neutrino masses given as $A=1.28335\times 10^{-3} \, \mathrm{eV}, B=1.24235\times 10^{-3} \, \mathrm{eV}$, $m_1=5.02563\times 10^{-2}\, \mathrm{eV},\, m_3=6.40312\times 10^{-3}\, \mathrm{eV}$. Thus, we find that the sum of neutrino masses takes the form $\sum m_\nu \sim 0.107659$ $\, \mathrm{eV}$,  which is consistent with the current cosmological constraints of Ref. \cite{constraint17}. This result is completely different from the one obtained in Ref. \cite{A4linear} where \emph{the inverted neutrino mass hierarchy is not allowed.}

\section*{Conclusion}
We have shown that some neutrino Yukawa terms must be added 
in the $A_4$ flavored linear seesaw model of Ref. \cite{A4linear}. When these terms are included, the obtained results are significantly different than the ones reported in Ref. \cite{A4linear}. Namely, in the normal neutrino mass spectrum, the leptonic mixing matrix in the case where the missed Yukawa terms are added is the same as in \cite{A4linear}, however, the expressions for neutrino masses are different from each other. On the other hand, in the inverted neutrino mass spectrum, our result is completely different from the one obtained in Ref. \cite{A4linear} and the sum of neutrino masses satisfies the current cosmological constraints, thus implying, unlike stated in Ref. \cite{A4linear} that this scenario is not ruled out.

\section*{Acknowledgments}
This research has been financially supported by Fondecyt (Chile), Grants
No.~1170803, CONICYT PIA/Basal FB0821, the Vietnam National Foundation for
Science and Technology Development (NAFOSTED) under grant number 103.01-2017.341.
H.N.L acknowledges the financial support of the  Vietnam Academy of Science and Technology under grant  NVCC05.01/19-19.

\end{document}